\documentclass[12pt]{iopart}

\usepackage{color}
\usepackage{cite}

\definecolor{DkRed}{cmyk}{0,.5,.5,.3}

\def\includefigs{y}
\newcommand{\cut}[1]{}

\newcommand{\resub}[1]{{\bf\color{DkRed}#1}}
\renewcommand{\resub}[1]{#1}
\usepackage{times}
\usepackage{graphicx}
\usepackage{verbatim} 
\usepackage{bm}

\begin{document}
\title[Diffusive hidden Markov model]{Diffusive hidden Markov model characterization of DNA looping dynamics in tethered particle experiments}
\author{John F Beausang and Philip C Nelson }


\address{Department of Physics and Astronomy, University of  Pennsylvania, Philadelphia, Pennsylvania 19104 USA, fax: (215) 898-2010, phone:(215) 898-7001 }
\ead{nelson@physics.upenn.edu}


\vspace{2pc}

\submitto{\PB}

\newcommand{\CO}{{\cal O}}
\newcommand{\CQ}{{\cal Q}}
\newcommand{\vr}{\mathbf{r}}
\newcommand{\steth}{\error}
\newcommand{\sloop}{_{{\rm loop}}}
\newcommand{\sun}{_{{\rm unl}}}
\newcommand{\sbead}{_{{\rm bead}}}
\newcommand{\sdiff}{_{\rm diff}}
\newcommand{\stot}{_{\rm tot}}
\newcommand{\Dmat}{\mathsf{T}}
\newcommand{\sdhmm}{_{\scriptscriptstyle\mathrm{DHMM}}}
\newcommand{\shmm}{_{\scriptscriptstyle\mathrm{HMM}}}

\newcommand{\tlf}{\tau_{\scriptscriptstyle\rm LF}}
\newcommand{\tlb}{\tau_{\scriptscriptstyle\rm LB}}
\newcommand{\transf}{T}
\newcommand{\gfunc}{G}

\def\max{_{{\rm max}}}
\def\rhomax{\rho_{\rm max}}
\def\kbt{k_{\rm B}T}
\def\mMunit{\ensuremath{\mbox{m\textsc m}}}
\def\nMunit{\ensuremath{\mbox{n\textsc m}}}
\def\that{\hat t}
\def\msunit{\ensuremath{\mbox{ms}}}
\def\pNunit{\ensuremath{\mbox{pN}}}
\def\bpunit{\ensuremath{\mbox{bp}}}
\def\nmunit{\ensuremath{\mbox{nm}}}
\def\umunit{\ensuremath{\mu\mbox{m}}}
\def\CO{{\cal O}} 
\def\CU{{\cal U}}
\def\CP{{\cal P}}
\def\sunit{\ensuremath{\mbox{s}}}
\def\half{{\textstyle\frac12}}

\newcommand{\pitemize}{\begin{itemize}\setlength{\itemsep}{0pt}\setlength{\parsep}{0pt}}
\newcommand{\xpitemize}{\end{itemize}}
\newcommand{\penumerate}{\begin{enumerate}\setlength{\itemsep}{0pt}\setlength{\parsep}{0pt}}
\newcommand{\xpenumerate}{\end{enumerate}}

\newcommand{\eff}{_{\rm eff}}

\newcommand{\capitem}[1]{(\textsf{#1})~}
\def\exv#1{\langle #1\rangle}
\def\dd{\mathrm{d}}
\def\ex#1{{\rm e}^{#1}}
\newcommand{\ee}[1]{\cdot10^{#1}}
\newcommand{\eem}[1]{\cdot10^{-#1}}
\newcommand{\pnvec}{\mathbf}
\newcommand{\inv}{^{\raise.15ex\hbox{${\scriptscriptstyle-}$}\kern-.05em 1}}

\def\pnlabel#1{
\label{#1}}
\def\eref#1{~(\ref{e:#1})}
\def\erefs#1{~(\ref{e:#1}} 
\def\sref#1{section~\ref{s:#1}}\def\Sref#1{Section~\ref{s:#1}}
\def\fref#1{figure~\ref{f:#1}}
\def\Fref#1{Figure~\ref{f:#1}}
\def\tref#1{table~\ref{t:#1}}
\def\frefs#1{figures~\ref{f:#1}}
\def\Frefs#1{Figures~\ref{f:#1}}
\def\frefn#1#2{\fref{#1}\panel{#2\/}}
\def\Frefn#1#2{\Fref{#1}\panel{#2\/}}

\def\frefsn#1#2{\frefs{#1}\panel{#2\/}}
\def\rref#1{Ref.~\cite{#1}}
\def\rrefs#1{Refs.~[\citen{#1}} 
\def\Rrefs#1{Refs.~\cite{#1}}

\def\aref#1{~\ref{a:#1}}
\def\Aref#1{~\ref{a:#1}}

\def\yesflag{y}
\def\ifig#1#2#3{\begin{figure}[tb!]
\begin{minipage}[b]{.95\textwidth}
\ifx\includefigs\yesflag
\begin{center}
\includegraphics
{Graphics/#3}\end{center}\fi
 \end{minipage}
\hfil\par\smallskip \caption{\footnotesize #2 \pnlabel{f:#1}}
\end{figure}%
}
\def\ifigab#1#2#3#4{
\begin{figure}[tb!]
\ifx\includefigs\yesflag%
\makebox[1.1\textwidth]{\begin{minipage}[b]{.45\textwidth}
\begin{center}\includegraphics{Graphics/#3}\end{center}
\end{minipage}\hfill
\begin{minipage}[b]{.45\textwidth}
\begin{center}\includegraphics{Graphics/#4}\end{center}
\end{minipage}\hfill}
\fi
\par\smallskip
\caption{\footnotesize #2 \pnlabel{f:#1}}
\end{figure}}


\maketitle
\begin{abstract}
%
%

In many biochemical processes, proteins bound to DNA at distant sites are brought into close proximity by loops in the underlying DNA. For example, the function of some gene-regulatory proteins depends on such ``DNA looping'' interactions. We present a new technique for characterizing the kinetics of loop formation \textit{in vitro}, as observed using the tethered particle method, and apply it to experimental data on looping induced by \textit{lambda} repressor. Our method uses a modified (``diffusive'') hidden Markov analysis that directly incorporates the Brownian motion of the observed tethered bead. We compare looping lifetimes found with our method (which we find are consistent over a range of sampling frequencies) to those obtained via the traditional threshold-crossing analysis (which can vary depending on how the raw data are filtered in the time domain). Our method does not involve any time filtering and can detect sudden changes in looping behavior. For example, we show how our method can identify transitions between long-lived, kinetically distinct states that would otherwise be difficult to discern.

\end{abstract}

\section{Introduction and summary}
\subsection{Tethered particle motion}
Molecular biophysics relies on an ever-expanding toolkit of methods to ``see'' processes not visible by traditional light microscopy. The toolkit is vast in part because each method has its own set of strengths and weaknesses, suiting it for a particular class of measurement challenges. For example, some methods may yield excellent spatial resolution but limited information about the dynamics of a process, particularly in a physiologically relevant context.

Recent methods that observe state transitions in single molecules allow us to see differences in behavior between different molecules, as well as making it possible to extract detailed kinetic information. Tethered particle motion (TPM) is one such technique to monitor conformational changes in single molecules of DNA, such as loop formation (\fref{cartoon}), in real time. Our goal in this paper is to extend the range of applicability of TPM as a tool to study the kinetics of such changes, adding many details to a previous letter \cite{beau07a}.

In TPM, a bead is tethered to the surface of a microscope slide by a single DNA molecule and undergoes Brownian motion without any externally applied force (for example, see
\cite{scha91a}--\nocite{yin94a,finz95a,gell95a,zocc01a,doho01a,vanz03a,sing03a,poug04a,toli04a,dixi05a,blum05a}\cite{wong07a}). Lateral displacements of the bead are passively observed through an optical microscope. TPM is suited for the study of short molecules (from 100 to several thousand basepairs long).
Unlike methods that stretch the DNA, TPM allows the study of proteins acting on molecules not subjected to externally applied tension. Thus TPM offers the possibility to quantify the dynamics of loop formation and breakdown, and then determine the dependence on biophysical parameters such as loop size, tether length, repressor concentration, operator sequence, and interoperator sequence. An additional advantage of TPM is that it can be parallelized: Many tethered particles can be simultaneously tracked in a single run.

\ifig{cartoon}{A DNA molecule flexibly links a bead to a surface. The motion of the bead's center is observed and tracked, for example as described in \rrefs{zurl06a,nels06a}. In each video frame, the position vector, usually projected to the $xy$ plane, is found. After drift subtraction, the mean of this position vector defines the anchoring point; the vector $\vr\cut{_{\perp}}$ discussed in this article is always understood to be the drift-subtracted, projected position, measured relative to this anchoring point. The length of this vector is called $\rho$. Time constants $\tlf$ and $\tlb$ determine the rates to form and break loops, respectively. The figure is simplified; in the actual experiment leading to the data we analyze, the DNA had two sets of three binding sites (``operators'') for cI repressor protein \cite{beau07a}, and each operator can at a given moment be occupied or unoccupied.}{cartoon.eps}

Some of the current implementations of TPM allow single-particle tracking of multiple beads with 20--50$\,\msunit$ time and $\sim$10 $\nmunit$ spatial resolution, allowing rather precise determination of effective tether length from data \cite{nels06a, sega06a,qian99a}. (Other current work observes a time-averaged image instead of using single particle tracking \cite{wong07a}.) Thus TPM provides an attractive assay for studying the formation of DNA loops by protein complexes that bind to multiple sites on a molecule, including the kinetics of such loop formation in solution. For example, van\thinspace{}den\thinspace{}Broek \textit{et al.} have performed such a measurement, using as their model system the restriction enzymes NaeI and NarI \cite{broe06a}. Other single-molecule techniques exist for observing juxtaposition of parts of molecules (e.g., F\"orster resonance energy transfer \cite{mcki06a}). These methods do not require the relatively large reporting particle (the bead) needed in TPM. But fluorescence techniques are subject to photobleaching, which limits their usefulness when studying very long-timescale transitions. In contrast, the TPM method easily allows the observation of a single DNA molecule over periods of 30--60 minutes \cite{zurl06a,nels06a}. Even for short-timescale transitions, the long observations possible with TPM give an advantage, because the hidden Markov analysis developed in this paper can extract reliable information from long time series, even when it has low signal to noise ratio.

Here we present an analysis method for TPM experiments on DNA looping that models the bead motion and looping transitions in terms of quantities determined empirically from data on non-looping DNA, plus two additional rate constants (for loop formation and breakdown). Then we apply a maximum-likelihood approach to determine the numerical values of the rate constants. We illustrate our approach by applying it to looping data from the \textit{lambda} regulatory-protein complex. In this context (and others such as the \textit{lac} complex \cite{finz95a,wong07a}), the analysis of looping kinetics can answer basic biological questions about gene regulation \cite{reve99a,dodd01a,hoch02a}. For example, is repression accomplished via a permanent loop? Or, does the loop open transiently, allowing some subthreshold activity of transcription in the ``repressed'' state? In the case of {lambda} phage, the latter scenario could allow faster response to changes in the cell's environment, without an unacceptable rate of ``accidental'' transitions to the lytic program. Indeed using our method, we found dynamic looping activity at concentrations of cI repressor protein corresponding physiologically to the lysogenic state of lambda  \cite{beau07a}.  Also, the cooperativity between multiple cI proteins bound to the \textit{lambda} DNA may result in loops of varying stability as operator occupancy changes on a time scale slower than individual looping events.

\subsection{Hidden Markov modeling}
The technical problems we wish to address are easily stated. We wish to deduce the kinetics of an invisible process (DNA loop formation) indirectly, by observing bead motion.  As loops form and break, the effective DNA tether length switches between two (or more) distinct values. Simply observing the magnitude of the bead's motion to determine the state of the tether (and thus the rate of loop formation) is not sufficient, because the expected distributions of projected bead position for the looped and unlooped states overlap substantially (\cite{sega06a}; see \fref{FIGHISTOS} below). We need a way to avoid misinterpreting these cases as looping events. We could try to minimize this problem by filtering the data over a large enough time window. For example, Refs.~\cite{broe06a, zurl06a} used a window of width 4$\,\sunit$, but such filtering degrades our ability to see brief but genuine looping events. Indeed we will argue that in some systems, for example the \textit{lambda} system studied in this paper, the mean state lifetimes are not much longer than the characteristic time scale for the bead to diffuse through its allowed domain. Windowing the data will then not yield unambiguous results; the kinetic rate constants inferred by the method of \cite{broe06a}, applied to our system, depend on the arbitrary choice of window size (see
\cite{beau07a,vanz06a} and \sref{tca} below).

Our difficulty is in some ways similar to that faced in the interpretation of ion-channel data, where again one wishes to infer invisible state changes from their visible, but noisy, consequences. In that context, hidden Markov modeling, combined with the maximum likelihood method (the ``HMM technique'' \cite{rabi89a}), offers advantages over simpler histogram schemes (see for example \cite{qin00a,qin00b}). The HMM technique supposes a particular kinetic scheme for the underlying process with unknown rate constants. For any given values of those constants, and any time series of the observable, one then calculates the likelihood  of that time series emerging in that model. Substituting actual experimental data for the time series, the inferred values for the unknown time constants are the ones that maximize this likelihood.

The HMM technique does not require time-domain filtering (windowing) to the data (although for some applications it may be desirable, e.g. \cite{smit01a,mile06a}). Nor does it require the selection of a threshold, across which each data point is labeled as being definitively in the looped or unlooped state. Finally, when the underlying process has more than two states, HMM in principle allows one to extract a more complete set of rate constants than does the method of dwell time histograms \cite{mile06a,mcki06a}.

For such reasons, HMM and other maximum-likelihood approaches have recently proven to be a powerful technique for the analysis of a number of single-molecule experiments [\citen{mcki06a,yang02a}--\nocite{yang03a,andr03a,schr03a,mile06b}\citen{hoog06a}]. Much of this paper will be devoted to adapting the HMM method to the context of TPM measurements of DNA looping. To accomplish this goal, we will need to overcome some obstacles, which we now outline.

\subsection{Need for diffusive HMM\pnlabel{s:ndhmm}}
TPM experiments generate data sets consisting of a set of bead images, from which  a time series of bead center locations $\CO=\{(x_{t},y_{t})\}=\{\mathbf{r}_{t}\}$ is determined \cite{nels06a}. Underlying this observed signal is the desired sequence $\{q_{t}\}$ of the DNA conformational state (e.g., in a $2$-state system, $q=1$ means unlooped, and $q=2$ looped, at each time point). In principle, the probability distribution $P(\vr_{t},q_{t})$ for the bead position and looping state at time $t$ could depend on the entire prior history of the system, that is, on $\mathbf{r}_{t-1},q_{t-1},\mathbf{r}_{t-2},q_{t-2}\ldots \mathbf{r}_{1},q_{1}$. Standard hidden Markov modeling \cite{rabi89a} assumes that the observed signal is uncorrelated, depending only on the current hidden state via a distribution $p\sbead(\mathbf{r}_{t}|q_{t})$, and that this hidden state in turn depends only on the previous one, via a ``transition matrix'' $\transf(q_{t}|q_{t-1})$:
\begin{equation}
P\shmm(\mathbf{r}_{t},q_{t})=p\sbead(\mathbf{r}_{t}|q_{t})\transf(q_{t}|q_{t-1}) \label{e:phmm}\end{equation} As we will show in \sref{tha}, however, this assumption does not apply to TPM experiments, in part because of correlations in position due to the diffusive character of the bead's motion. Essentially, our modified DHMM replaces the above assumption with a slightly more general version, in which $P(\vr_{t},q_{t})$ depends on \textit{both} $\vr_{t-1}$ and $q_{t-1}$.  Finding and justifying a form for this probability function, with just two unknown fit parameters, is the crux of our problem.

\subsection{Outline of this paper and summary of notation}

Table~\ref{t:notation} in the Glossary section gives a summary of the notation used in this paper. Readers familiar with TPM experiments and hidden Markov modeling may wish to use this table and skip directly to \sref{dm}, where we formulate
our approach.

\Sref{te} outlines the class of experiments we study and shows some typical data. \Sref{pw} outlines methods used in prior work, specifically the windowing/threshold method and standard hidden Markov modeling, along with our critique. \Sref{r} gives some illustrative results for the experimental dataset studied in \rref{beau07a}.

\section{Typical experiments\pnlabel{s:te}}
\ifig{FIGTIMERHO}{Typical time series of bead positions. DNA constructs of total length $3477\,\bpunit$ were attached at one end to a glass coverslip and at the other to a $480\,\nmunit$ diameter bead. The vertical axis gives the actual distance of the bead center from its attachment point, after drift subtraction. The trace shown passed the tests discussed in \rref{nels06a}, for example the ones that eliminate doubly-tethered beads. The DNA construct contained two sets of three operator sites. The two sets of operators were separated by $2317\,\bpunit$. The system contained cI repressor protein at concentration 200$\,\nMunit$ (where 200$\,\nMunit$ refers to the concentration of cI dimers since monomers do not bind to DNA); repressor proteins bind to the operator sites on the DNA, and to each other, looping the DNA as in \fref{cartoon}.  A sharp transition can be seen from a regime of no loop formation to one of dynamical loop formation at $\sim650\, \sunit$. Later \sref{dvllst} will argue that the latter regime itself consists of two kinetically distinct subregimes. The dashed lines represent the two values of $\rhomax$ corresponding respectively to the looped and unlooped states; control data in these two states was observed never to exceed these values. A brief sticking event, indicated by the inverted triangle, was excised from the data prior to analysis. \Sref{osdf} uses these observed values of $\rhomax$ to create truncated Gaussian step distributions for our models of the looped and unlooped Brownian motion. (Experimental data for this and subsequent figures from \cite{beau07a}, kindly supplied by C.~Zurla and L.~Finzi.) }{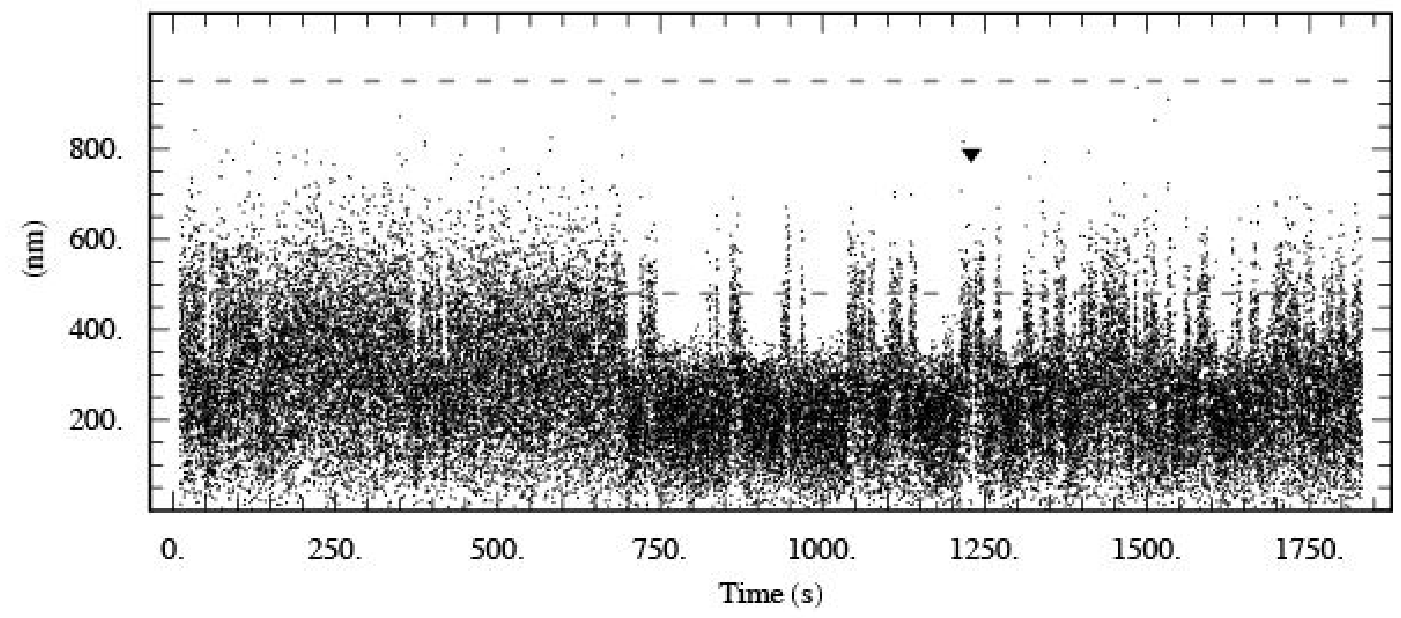}

\ifigab{FIGTIMEXY}{Time series for $x$ and $y$ corresponding to \fref{FIGTIMERHO}. The graphs  show that our drift subtraction scheme leads to visually similar traces for $x$ and $y$.}{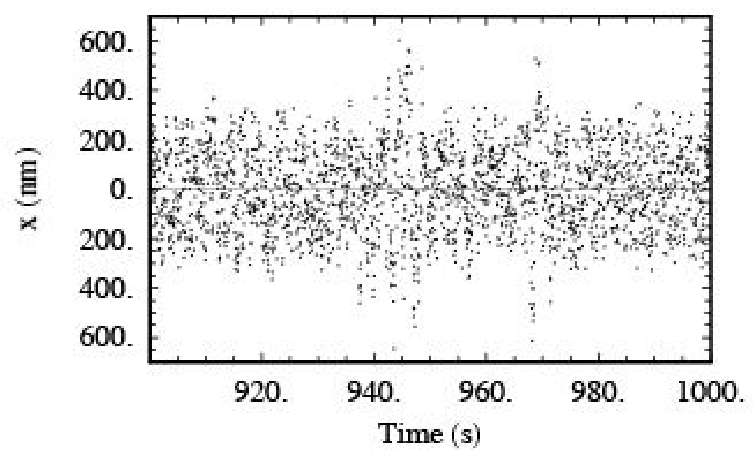}{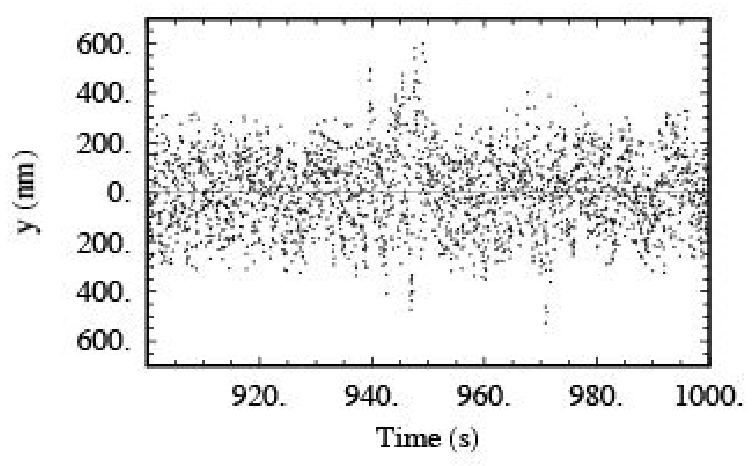}
For concreteness, we will analyze a representative TPM experiment \cite{zurl06a,nels06a,beau07a} in which the projected $(x,y)$ positions of up to 6 beads are simultaneously imaged using differential interference contrast microscopy and recorded along with a time stamp for up to 45 minutes. Images are recorded from alternate rows of pixels every 20$\,\msunit$ from a charge coupled device operating in interlaced mode. Due to the difficulty in obtaining precise alignment of the two rows of pixels, we treat each time series as two separate sets; the resulting 40$\,\msunit$ time resolution is adequate for our purposes. Rejection of anomalous beads (e.g., double tethers, surface adhesion, etc.) and correction for microscope drift are described in \rref{nels06a}. Simultaneous tracking of multiple beads allows microscope drift to be estimated from the collective motion of all the beads and then subtracted from each bead separately, \cite{nels06a}.  We determine the long--time drift by first finding the average $(x,y)$ position of each bead in a 20\,\sunit\, and then a spatial average over all the beads is calculated by summing over each of the 20\,\sunit\, time averages.  An interpolating function that passes through these points is fitted and subtracted from each measurement \cite{nels06a}. This second average over the beads results in a smoother interpolation that minimizes any inadvertent removal of true bead motion.  Rarely (2 points in \fref{FIGTIMERHO}), an anomalous outlying point or a missing frame is replaced with an interpolation of the neighboring points.  Whenever we refer to $x$ and $y$ below, we are referring to the drift-corrected time series.

\ifig{FIGHISTOS}{\textit{Left solid curve:} Normalized probability distribution function (pdf) of bead center location for experimental control data corresponding to a permanently looped tether (about $65\,000$ video frames). \textit{Right solid curve:} Corresponding pdf for unlooped control data (about $200\,000$ video frames). \textit{Dashed curves:} Corresponding pdfs for simulated bead motion, computed using the step distribution functions found in \sref{osdf}, with similar numbers of simulated steps. Although the agreement with the experimental data is not perfect, it is quite nontrivial: The step distribution functions were not chosen to make these distributions agree, but rather to match the observed distributions of steps between pairs of adjacent video frames. In contrast to this dynamical model, the equilibrium calculations of \rrefs{sega06a} and \cite{nels06a} are less ambitious; they evaluate expectations based on Monte Carlo integration of the Boltzmann distribution. (Data from \cite{beau07a}.) }{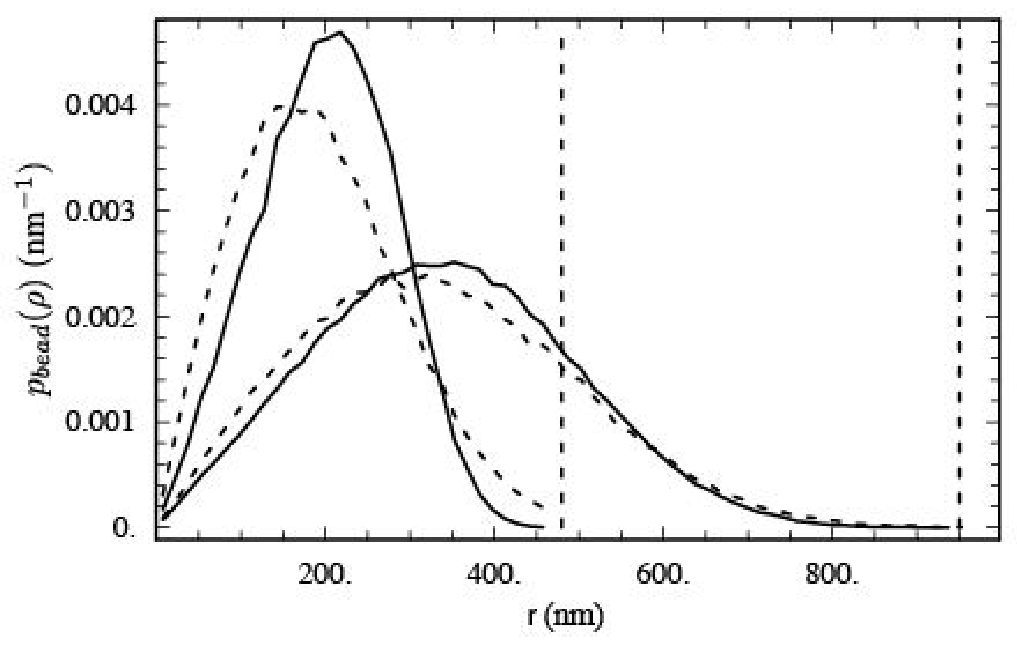}

\ifig{FIGAUTO}{\textit{Upper solid curve:} Semilog plot of the time autocorrelation function, $\half\langle\vr_{t}\cdot\vr_{t+\tau}\rangle$ for unlooped experimental control data used in \fref{FIGHISTOS}. \textit{Lower solid curve:} Analogous function for permanently looped control data. \textit{Dashed curves:} Analogous functions for simulated control data, computed using the step distribution functions found in \sref{osdf}. As in \fref{FIGHISTOS}, the agreement is nontrivial: The simulation was based on pairs of data points differing by just one video frame, so the approximate agreement supports our assumption that the tethered bead motion is Markovian. (The equilibrium calculations of \rrefs{nels06a},\citen{sega06a}] made no attempt to calculate autocorrelations.)} {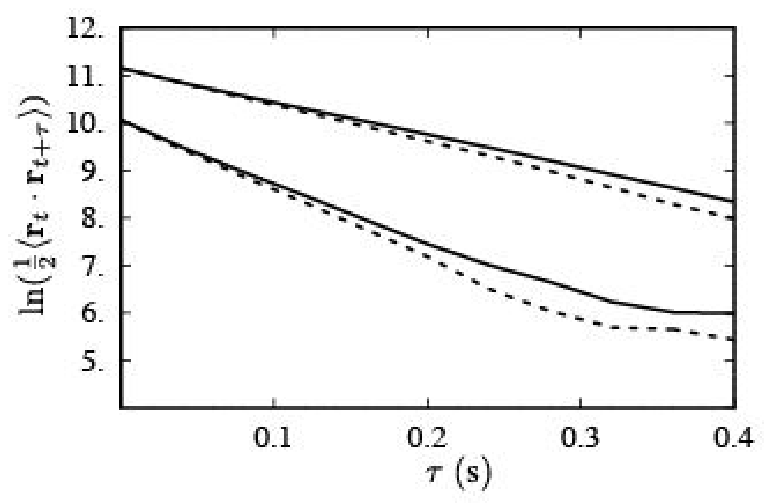}

Adding 200$\,\nMunit$ of cI repressor protein converts the homogeneous tethered Brownian motion of the particle to a regime characterized by abrupt, dynamic transitions in bead motion (\frefs{FIGTIMERHO}--\ref{f:FIGTIMEXY}), a fraction of which have  long enough dwell times ($\sim$1--100 sec) to be visible by eye in the microscope.

We also use data from two kinds of control experiments: \textit{(1)}~tethered beads with no protein and thus no looping, and \textit{(2)} a small subset of beads with cI present that remain in a permanently looped configuration for many minutes \resub{(data not shown, see \cite{beau07a})}. The first of these controls is obtained with every experiment from a preliminary 20 minute recording with no repressor present. \Frefs{FIGHISTOS}--\ref{f:FIGAUTO} show two important reduced forms of typical experimental data (solid curves), along with simulation results to be discussed in later sections.

\section{Prior methods\pnlabel{s:pw}}
Before constructing the DHMM method, we briefly review two existing methods, to highlight their limitations in the context of tethered particle motion experiments.

\subsection{Windowing/threshold method\pnlabel{s:tca}}
A traditional method for determining transition rates from a signal that depends on the hidden, underlying state is to record the signal over many transitions; transition events are defined when the signal crosses some threshold value.  The dwell times between threshold crossings of the experimentally distinguishable states are separately histogrammed and then fit to an exponential (or multiexponential) distribution. The parameters of the fits are the state lifetimes; their inverses are the transition rates.

TPM experiments present some difficulties for this straightforward approach, however. First, the observed signal (bead center position) is only indirectly related to the desired DNA looping state. For example, in the unlooped state the bead will spend an appreciable fraction of its time close to the attachment point, mimicking the looped state, see \fref{FIGHISTOS}. \cut{; in other words the probability distribution functions (pdfs) of bead position in the two states have considerable overlap. }Additionally, the height coordinate $z$ of bead position is either not measured (e.g.\ in \cite{zurl06a,nels06a}) or is measured to less precision than $x,y$ (e.g.\ in \cite{blum05a}), increasing further the overlap between the pdfs (\fref{FIGHISTOS}). Thus we cannot state with certainty the looping states of a majority of the individual data points in our time series; in our example data, measurements with $\rho_{t}$ less than  $\sim 450\,\nmunit$ can be in either looped state.

To overcome this difficulty, the raw data are often filtered using a sliding time window of width $W$, whose value is chosen to make clearly visible steps emerge in the data. Often this filtering step is a calculation of the signal's variance in the window, which has the additional advantage of being insensitive to instrumental drift over time scales slower than $W$. Next, a threshold is chosen that separates the high- and low-variance states, which now appear more clearly in the filtered data. Dwell times are then defined between those threshold crossings of the filtered data whose durations exceed the filter dead time (see \cite{colq83a,vanz06a}). If the window $W$ contains many independent measurements, then we get an accurate measure of the variance, and hence few spurious reported threshold crossings between the high- and low-variance states.

Unfortunately, any filter based on a finite-sample estimate of bead position variance converges slowly as we increase $W$. For a time series of \textit{uncorrelated} samples, the variance of the estimated variance is proportional to $1/(\mbox{number of samples in the window }W)$ \cite{bevi03a,smit01a}. This observation may make it seem that we need only take data at a high enough frame rate $\gamma$ to make $1/(W\gamma)$ sufficiently small. But successive snapshots of bead position are \textit{not} uncorrelated. The appropriate prefactor is not $1/(W\gamma)$ but rather $\tau\sdiff/W$, where $\tau\sdiff$ is the diffusion time for the bead to traverse its range of motion. We estimate $\tau\sdiff\approx 140\, \msunit$ for TPM experiments from the $1/e$ decay of the position autocorrelation function with $\sim$\umunit\ length lambda DNA fragment and $\sim0.5\,\umunit$ diameter beads.

Hence the fractional statistical noise in our estimate of bead variance, $\sqrt{\tau\sdiff/W}$, decreases slowly with increasing $W$; $W$ must therefore be taken large to obtain sufficient noise rejection (sufficiently rare spurious threshold crossings). Indeed, we found using control data that taking $W$ less than 1--2 seconds introduces spurious looping events in the unlooped control data where there should be none. On the other hand, $W$ must be taken smaller than the shortest state lifetime ($\sim1\,$second); otherwise we miss too many genuine state transitions. These two constraints set a fundamental limit on the accuracy of the windowing/threshold method, and in fact we found that applying the threshold-crossing method to experimental data of reasonable duration resulted in reported lifetimes that depend on the choice of $W$ (see \fref{FIGTHRESHBOTH}) \cite{beau07a}.

\ifig{FIGTHRESHBOTH}{\textit{Black dots:} Lifetime results for the windowing/threshold method for DNA loop formation lifetime (points), as a function of filter window size $W$. These results are based on the segment of data between 650--1750 seconds in \fref{FIGTIMERHO} (total of $27\,500$ video frames). For each value of $W$,  dwell times longer than twice the filter dead time \resub{(i.e. $W$)} were histogrammed and fit to single exponentials to determine \resub{$\tlf$}. No attempt was made to correct for missed events. \textit{Open dots: }Similar results for loop breakdown time scale \resub{$\tlb$}. \resub{The dependence of the inferred lifetime on the filter can complicate the ``best'' choice of $W$.} Our DHMM method uses no window.}{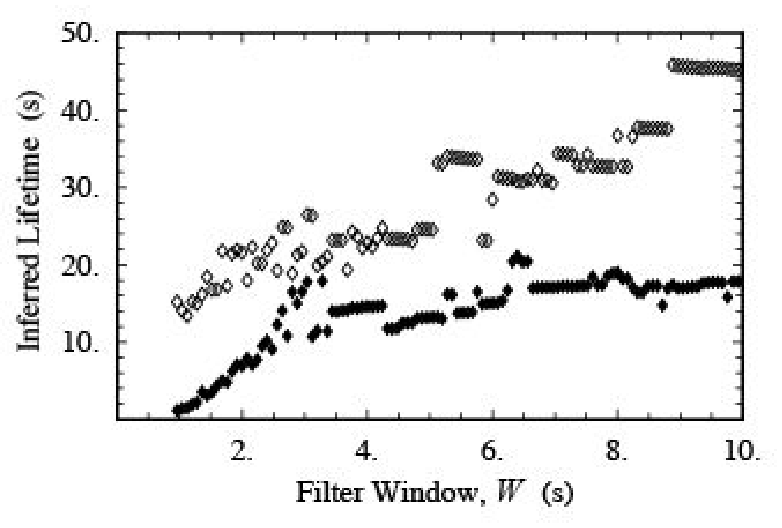}

Techniques do exist to correct for missed events \cite{colq83a,broe06a,vanz06a} and to improve the reliability of the threshold method.  Our DHMM model will bypass this difficulty altogether, by avoiding the filtering step.

\subsection{Standard HMM analysis\pnlabel{s:tha}}
This section briefly reviews the standard hidden Markov method, considers its possible application to tethered particle experiments, and points out a serious limitation in its applicability. The purpose is to motivate our modification of standard HMM in \sref{dm}, and to establish some mathematical notation to be used there.

DNA looping experiments are natural candidates for application of hidden Markov methods, because the probability distribution functions (pdfs) for the bead position (called $p\sbead(\mathbf{r}|q)$ in \eref{phmm}) can be calculated \textit{a priori} \cite{sega06a,nels06a} or measured directly \cite{zurl06a,nels06a}. To model dynamical looping data, then, we could imagine an underlying (hidden) 2-state Markov process describing loop formation and breakdown, then suppose that in each of the two states the observed quantity (bead position) is drawn from the pdf appropriate to the current looping state. The underlying process has two unknown parameters, the rate constants $1/\tlf$ for loop formation and $1/\tlb$ for loop breakdown. Using assumed values for these parameters, and the known pdfs for bead position, we can calculate the likelihood that any given time series of bead positions would be observed. (For a general introduction to maximum likelihood methods see \rrefs{bevi03a,cox74a}.) Substituting an actual observed time series and maximizing the likelihood over $\tlf$ and $\tlb$ then yields the best estimates for the lifetimes, which are the quantities of interest to us.

Before we critique the approach just outlined, let us make it somewhat more explicit. The standard HMM \cite{rabi89a} supposes that an observed signal reflects two processes: An autonomous Markov process (in our case, loop formation and breakdown) generates a time series $\CQ =\{q_{t}\}$ at a discrete set of times $t=1,\ldots M$, using a $2\times2$ matrix of transition probabilities to represent $\transf (q|q')$ in \eref{phmm}, where the $t$ subscripts are dropped and the prime denotes the preceding measurement. $\CQ $ is not directly observable, but it influences an observable signal $\CO =\{\rho_{t}\}$: At each time, $\rho_{t}$ is drawn from a probability distribution $p\sbead(\rho_{t}|q_{t})$, which depends only on the current value of $q_{t}$. In particular, there is assumed to be no ``back reaction'' from the observed value of $\rho$ onto the underlying Markov process, and no ``memory'' in the process that generates $\CO=\{\rho_{t}\}$.

The overall likelihood of observing a time series $\CO $ is then taken to be a sum over all possible hidden trajectories $\CQ $: \begin{equation} P_{\rm tot,\scriptscriptstyle HMM} (\CO)=\sum_{\CQ }\left[\prod_{t=2}^{M} p\sbead(\rho_{t}|q_{t})\transf(q_{t}|q_{t-1})\right] \pi(\rho_1,q_{1}) \pnlabel{e:ptothmm}\end{equation} where we initialized the product with the probabilities $\pi(\rho_1,q_{1})$. In our case,  $\rho$ is the radial distance $\rho=\sqrt{x^2+y^2}$ from the projected bead center to the tether attachment point (because $p\sbead$ is circularly symmetric).

The evaluation of \eref{ptothmm} may at first seem prohibitively difficult: Typically we may have $M=70\,000$ video frames, and hence $2^M$ terms in the sum over $\CQ $! But closer inspection shows that \eref{ptothmm} is the product of $M$ $2\times2$ matrices, which can be evaluated in order-$M$ steps. Namely, \eref{ptothmm} can be rewritten as
\begin{equation}
P_{\rm tot,\scriptscriptstyle HMM} (\CO) =(1,1)\cdot \Dmat\shmm(\rho_{M})\cdots\Dmat\shmm(\rho_{2})\Pi(\rho_1) \pnlabel{e:ptothmmMat}\end{equation}
where
\begin{equation}
\Dmat\shmm (\rho)= \left(
\begin{array}{cc}1-(\Delta t/\tlf)&(\Delta t/\tlb)\\
(\Delta t/\tlf)&1-(\Delta t/\tlb)\\\end{array} \right) \left(\begin{array}{cc}
p\sbead(\rho|q = 1)&0\\
0&p\sbead(\rho|q=2)\\ \end{array}\right) \pnlabel{e:dmathmm}\end{equation}
The row vector $(1,1)$ is used in \eref{ptothmmMat} to obtain a scalar probability, and $\Pi(\rho)_q=\pi(\rho,q)$ is regarded as a column vector. The first factor in \eref{dmathmm} is just $\transf(q|q')$ regarded as a matrix; $\Delta t$ is a time step that is much smaller than $\tlf$ or $\tlb $.

With this insight, evaluating the expression and even optimizing it over the parameters $\tlf $ and $\tlb $ is not so prohibitive after all.  A numerical issue does arise when dealing with long time traces, in that the value of $P\stot$ becomes very small.  \resub{To handle this, we perform a normalization procedure after each time step \cite{rabi89a} in the calculation of \eref{ptothmmMat}.  For example, after the first $t$ multiplications we divide each of the two entries in the matrix by their sum $s_t$, thus keeping all of the terms in the multiplication  close to unity. We maintain a running tally $S_t$ of the normalizing factors $\ln s_t$ so that by the end of the calculation of \eref{ptothmmMat}, the final value $S_M$ is the logarithm of the desired $P\stot(\CO)$.}

One problem with the procedure outlined above, as mentioned in \sref{ndhmm}, is that it neglects correlations in observed bead position due to the Brownian character of tethered particle motion. Equation \eref{ptothmm} assumes that the ``noise'' (bead motion) has no memory. Although this may be a valid assumption for the noise in ion channels, it is certainly not the case for tethered particle motion, if the sample times are separated by less than the bead's diffusion time. And indeed, we found that HMM in this form was not able to determine reasonable lifetimes in experimental data (\fref{FIGHMMPLATEAU}). Instead, the algorithm reports very short lifetimes, inconsistent with obvious looping events visible in the data. This spurious short lifetime depends on the sampling rate of the measurement, and is present in data from control experiments with no looping.

\Aref{a} illustrates an \textit{ad hoc} approach for addressing bead diffusion in an HMM analysis by thinning the data. While this may be acceptable in some circumstances, we found that a better approach is to modify the HMM procedure itself to include bead diffusion directly. \Sref{dm} will show that the required modification is simple, and computationally no more difficult than the calculations sketched in this section.

\section{DHMM method\pnlabel{s:dm}}
\subsection{Formulation\pnlabel{s:f}}
The previous section argued that the standard HMM approach cannot be applied to tethered particle motion because its starting point, \eref{ptothmm}, is not valid. First, even if the rate of loop formation is zero (as it is in the absence of any cI protein), there will still be correlations in the observed bead positions that the algorithm interprets as spurious transitions. In fact, the time series of bead positions itself has a Markov character: Each position $\vr_t$ is drawn from a probability distribution $\transf\sun(\mathbf{r}_{t}|\mathbf{r}_{t-1})$ depending only on $\vr_{t-1}$, independent of the positions at prior times \cite{beau07b}.

A second violation of \eref{ptothmm} is apparent when we note that the HMM assumption that the loop formation/breakdown process is autonomous (independent of the observed bead positions) is also not valid. As an extreme example, we note that if at some time the bead position is observed to be so far from the attachment point that the tether is stretched nearly straight, then it is geometrically impossible for a loop to form and the momentary rate of loop formation must equal zero.

In short, the dynamics of the bead/tether system must be regarded as a single, extended Markov process, with a joint conditional probability $\Dmat\sdhmm(\vr_{t}|\vr_{t-1})_{q_t,q_{t-1}}$; we replace \eref{ptothmmMat} by
\begin{equation}
P_{\rm tot,\scriptscriptstyle DHMM} (\CO)=(1,1)\cdot \Dmat\sdhmm(\vr_{M}|\vr_{M-1})\cdots\Dmat\sdhmm(\vr_{2}|\vr_{1})\Pi(\vr_1) \pnlabel{e:ptotdhmmMat}
\end{equation}
Henceforth we abbreviate $P_{\rm tot,\scriptscriptstyle DHMM}$ as $P_{\rm tot}$.

Note that since we wish to capture the dynamics of the bead motion, we can no longer reduce our description of the bead center from its position vector $\mathbf{r}$ to its length $\rho$, because there are pairs of points with the same $\rho$ that are nevertheless spatially distant. Replacing $\vr$ by $\rho$ would therefore discard some valuable, observed, information. We must, however, reduce from the full 3-dimensional position to its 2D projection when dealing with experiments in which $z$ is not observed. In this paper we will always make this reduction, so we simply write $\vr$ for the projected position.

What should we now take for our transition matrix $\Dmat \sdhmm(\mathbf{r}_{t}|\mathbf{r}_{t-1})_{q_t,q_{t-1}}$? It may seem that we have lost the most desirable feature of HMM, namely that the transition matrix was fully determined by the two fit parameters $\tlf$ and $\tlb$, along with functions known \textit{a priori} (the two empirically observed pdfs $p\sbead(\rho|q)$). We need a correspondingly simple proposal for the matrix $\Dmat \sdhmm $, which depends on  \textit{two} continuous quantities $\vr,\vr'$.

Our answer to this question is a heuristic proposal for $\Dmat\sdhmm $. Although not rigorously justified, our proposal retains the properties of depending on empirically determined functions and two fit parameters $\tlf,\tlb$; moreover, it incorporates the diffusive character of bead motion, which was not included in \eref{ptothmm}. The empirical functions describing the tethered motion of the bead are $\transf\sun(\mathbf{r}|\mathbf{r}')$ and $\transf\sloop(\mathbf{r}|\mathbf{r}')$ for the unlooped and permanently looped states, respectively. We first state our proposal, then discuss its meaning. Subsequent sections will show how to extract the empirical functions from control data, then how to apply the model to data on dynamic looping.

We assume that the matrix $\Dmat \sdhmm $ takes the form
\begin{equation}
\Dmat \sdhmm (\mathbf{r}|\mathbf{r}')= \left(
\begin{array}{cc}1-(\Delta t/\tlf)\Theta&(\Delta t/\tlb)\\
(\Delta t/\tlf)\Theta&1-(\Delta t/\tlb)\\\end{array} \right) \left(\begin{array}{cc}
\transf\sun(\mathbf{r}|\mathbf{r}')&0\\
0&\transf\sloop(\mathbf{r}|\mathbf{r}')\\ \end{array}\right) \pnlabel{e:dmatdhmm}\end{equation} In this formula, $\Theta\equiv 1-\Theta(\rho-\rhomax)$ is a step function, equal to 1 if looping is geometrically permitted, and 0 otherwise. The value of $\rhomax$ is to be obtained from experimental data, as described in \sref{i}.

When looping is either forbidden or obligatory, \eref{dmatdhmm} reduces to tethered Brownian motion in either the unlooped or looped states respectively (no dynamic looping). In other cases, we can think of the formula as describing an alternating series of transitions. First, the bead takes a diffusive step based on its current looping state (second matrix factor in \eref{dmatdhmm}). Next, the tether updates its looping state, using probabilities that depend in a simple way on its current position (first matrix factor in \eref{dmatdhmm}). Then the process repeats. If $\Delta t$ is much smaller than either the diffusion time or the loop formation/breakdown times, then we make no significant error by decomposing the process in this way. Note that \eref{dmatdhmm} is properly normalized, as we can check by summing it over all final states $\vr,q$.

\subsection{Critique of DHMM approach\pnlabel{s:c}}
Although it is reasonable, our proposal does make some strong and perhaps naive assumptions about the looping process. Before turning to the implementation of the method, we now pause to make some of these assumptions explicit, and indicate how future experiments could help justify them. The material in this section will not be used later; the reader may wish to skip to \sref{i}.

DNA loop formation involves the motion of the molecule through a high-dimensional space of shapes, driven by thermal motion, subject to a free energy landscape determined by the molecule's elasticity. When binding sites on the proteins encounter each other, or the DNA's operator sequences, binding may ensue depending on the precision of their alignment and the relevant binding constants. Phrased in this way, it's clear that the calculation of DNA loop formation kinetics is very complicated. However, such \textit{ab initio} calculations are not our goal.

Our goal is to develop a simple characterization for the looping behavior seen in TPM experiments, in a way that is also relevant for looping behavior \textit{in vivo}, and that also lets us observe differences in behavior as we change system parameters. We imagine that, for a free DNA chain, polymer dynamics repeatedly brings binding sites into juxtaposition at some rate, with a certain probability that any such encounter leads to formation of a bound state. The product of the attempt rate and the probability per encounter is an average loop formation rate. If we suppose that each encounter's probability for binding is small, then it is reasonable to expect that overall loop formation (and breakdown) processes should be monomolecular reactions describable with simple exponential kinetics.

Turning from free DNA to the case of DNA tethering a large reporter bead, we first note that the presence of the bead does not by itself alter the thermal force fluctuations on the looping part of the DNA; these equilibrium fluctuations are determined by equipartition applied to the entropic elasticity of the semiflexible polymer chain. It is true that bead-surface repulsion can tend to stretch the DNA, altering the equilibrium constant for loop formation \cite{sega06a}. However, this entropic force falls off rapidly when the distance from the polymer's endpoint to the surface exceeds the bead diameter; it can be minimized by choosing small enough beads (or by replacing the bead by a functionalized colloidal gold particle \cite{dunn07a}).

Notwithstanding the above remarks about equilibrium, we should expect that the large, sluggish bead will significantly alter looping \textit{kinetics}. But precisely because the bead is slowly diffusing, whenever it is close to the attachment point of the DNA to the wall, it is likely to stay close for many milliseconds. During this time, the bead offers no significant obstacle to the same thermally-driven chain rearrangements that bring the operator sites of free DNA into juxtaposition. Hence, we may expect that when the bead is close to the wall attachment point, DNA loop formation will proceed as if the DNA were free in solution (or part of a larger bacterial genome)\footnote{More precisely, we are assuming that the time required for the polymer tether to explore its available conformations is much less than the time for the bead to diffuse a significant fraction of its total range of motion.}. In the contrary case (the bead is far from the attachment point), loop formation is geometrically forbidden. A similar argument suggests that loop breakdown kinetics should not be altered by the presence of the reporter bead.

Equation\eref{dmatdhmm} embodies the above ideas, together with the idea that the bead wanders in and out of the range allowed for looping, subject to distributions $\transf\sloop$ and $\transf\sun$ that themselves can be found from the observed behavior of permanently looped or unlooped tethers. Thus we expect that the parameters $\tlf$ and $\tlb$ appearing in \eref{dmatdhmm} will, when the model is fit to TPM data, also give a good guide to looping rates for DNA free in solution. In contrast, simply applying the windowing/threshold method to data does not correct for the expected slowing-down of loop formation due to the presence of the bead. Indeed, applying that method to our data leads to a inferred $\tlf$ values significantly slower than the one we will obtain in \sref{r} below.

Although the preceding paragraphs have argued that our approach is reasonable, certainly it is somewhat crude. For example, once the bead center is observed to be at a certain distance from the attachment point, this distance amounts to a stretching of the DNA chain. The expected rate for loop formation will be some decreasing function of this stretching, but not of course a step function, as we assumed in \eref{dmatdhmm}.

Another simplification we have made is to ignore the unobserved height variable $z$, in effect treating the bead motion as diffusion in two dimensions. Although in \textit{free} Brownian motion all three coordinates perform independent random walks, in our problem the presence of the wall and tether couple $x,y,$ and $z$. To some extent, our technique of obtaining $\transf\sun$ and $\transf\sloop$ in \eref{dmatdhmm} directly from control data will correct for this effect, but the criterion for loop formation to be possible really depends on the full 3D separation $\sqrt{x^2+y^2+z^2}$, not on $\rho=\sqrt{x^2+y^2}$ as assumed in \eref{dmatdhmm}. A better analysis might treat $z$ as another hidden (unobserved) variable.

We also ignore rotatory Brownian motion of the bead. The ability of the tether to form loops actually depends on the distance between the two DNA end points. One endpoint, where the DNA attaches to the microscope slide, is fixed. We are using the bead center location as a proxy for the DNA-bead attachment point, but really the latter depends on both the former and also on the angular orientation of the bead. Again, a better analysis might treat this orientation as another hidden variable.

Our justifications for all three of the above simplifications are simply that \textit{(a)}~although the pdfs for projected bead position in the looped and unlooped states overlap partially, they are nevertheless fairly distinct, allowing reliable state identification even when we simplify some of the details in the model; \textit{(b)}~the projected-step distribution functions $\transf\sun$ and $\transf\sloop$ that we extract from control-experiment data do have the qualitative form that we would expect for two-dimensional diffusion in an effective spring trap \cite{beau07b} (see \sref{i}); and \textit{(c)}~changing our choice of the cutoff $\rhomax$ in the analysis does not significantly change our inferred values of the rate constants (data not shown). Despite these encouraging observations, however, other experimental tests of the method would certainly be desirable, for example, analyzing the kinetics of loop formation in identical tethers attached to different-sized beads, to check that similar values of $\tlf$ and $\tlb$ emerge. It may also be desirable to refine our mathematical model to account for some of the above issues; we leave these refinements to future work.

\subsection{Implementation\pnlabel{s:i}}
\subsubsection{Obtaining step distribution functions\pnlabel{s:osdf}}
Even granting the simplifications proposed in the preceding subsections, it would be a daunting task to determine the appropriate step distribution functions $\transf\sun$ and $\transf\sloop$ appearing in \eref{dmatdhmm} \textit{a priori} (directly from theory). For example, bead--wall hydrodynamic interactions depend on the bead's height, which is not observed; the tether couples the unobserved bead orientational fluctuations to the observed position fluctuations; and so on. We circumvent these difficulties by empirically determining the $\transf\sun$ and $\transf\sloop$ from experimental control data for the two states. After these functions are obtained, we confirm that the simple model of tethered-particle dynamics we construct from them indeed reproduces some nontrivial features of the real control data (\frefs{FIGHISTOS} and \ref{f:FIGAUTO}). Finally we examine dynamic-looping data, and adjust the two remaining free parameters $\tlb $ and $\tlf $ in \eref{dmatdhmm} until the log-likelihood, $\ln[P\stot(\CO)]$, is a maximum.

In order to obtain $\transf\sun(\mathbf{r}|\mathbf{r}')$ from the unlooped control data, we first note that this function must be symmetric under rotations of both $\mathbf{r}$ and $\mathbf{r}'$ about the attachment point by a common angle. Thus we only need to find this function for $\vr'$ on the $\hat\mathbf x$-axis, at some radial distance $\rho'$. Starting from a time series for unlooped DNA (no repressor protein present), we thus select all of the points in the time series for which the bead center's distance from the anchor point, $\rho'$, lies in a particular range. Next, we find the rotation in the plane that brings $\vr'$ to the $\hat\mathbf x$-axis, and apply that rotation also to $\Delta\mathbf{r}=\vr-\vr'$, the bead's vector displacement to its position on the following video frame. Finally, we build up a 2-dimensional histogram of the observed displacements $\Delta\mathbf{r}$, and normalize to obtain $\transf\sun(\vr|\vr')$. We then repeat the process, producing histograms for all observed initial distances $\rho'$. Using data obtained from about 30 minutes of bead observation, we found that we could divide the observed range of $\rho'$ into 30 intervals and still have reasonable statistics in the histograms; \frefs{FIGHISTO3D}--\ref{f:FIGHISTO3DPROJX} shows typical examples for two values of $\rho'$. We applied a similar procedure to the permanently-looped control data to obtain $\transf\sloop(\mathbf{r}|\mathbf{r}')$.

\ifigab{FIGHISTO3D}{2D histograms of unlooped control data at two values of $\rho'$ near (\textit{left}) and far (\textit{right}) from the anchor point (0,0) after rotation onto the $\hat{\bf x}$ and $\hat{\bf y}$ axes (see text). The dots on the upper x-y plane indicate the initial position $\rho'$ (\textit{full circle}) and the mean midpoint of the final position $\mu(\rho')$ (\textit{open circle}).  For $\rho'$ near the anchor point the two dots coincide.}{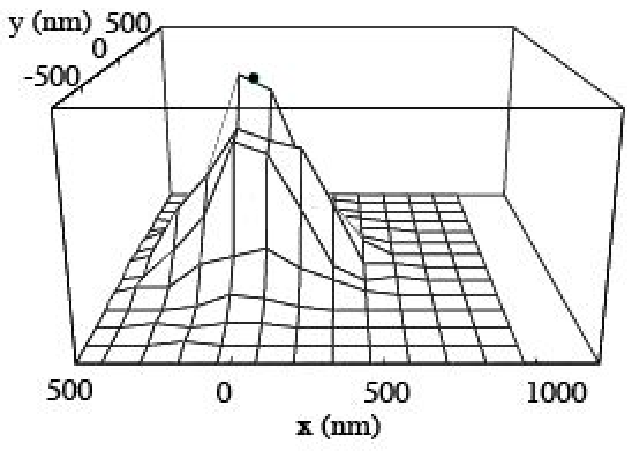}{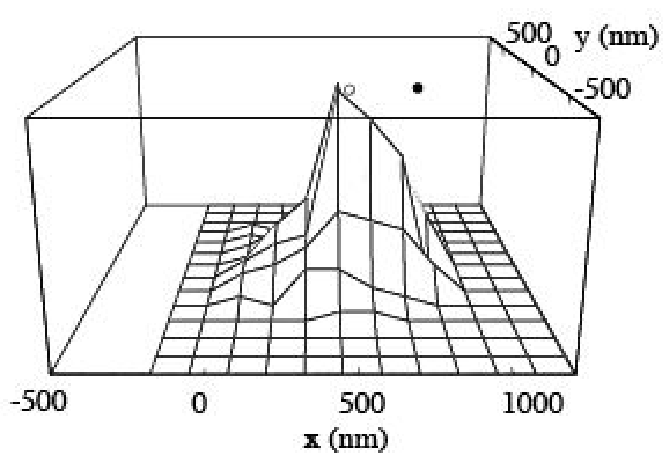}

\ifigab{FIGHISTO3DPROJX}{Projections of the two distributions in \fref{FIGHISTO3D} onto the $\hat{\bf x}$ and $\hat{\bf y}$ axes, together with Gaussian distributions chosen to idealize them. \textit{Left:} the \textit{vertical lines} represent two choices for the initial bead position; \textit{dots} represent the corresponding distributions of bead positions on the next video frame. Note the shift in the mean $\hat{\bf x}$ at larger $\rho'$ (\textit{open}) compared to shorter $\rho'$ (\textit{full}). \textit{Right:} No such shift is observed in the $\hat{\bf y}$ direction.}{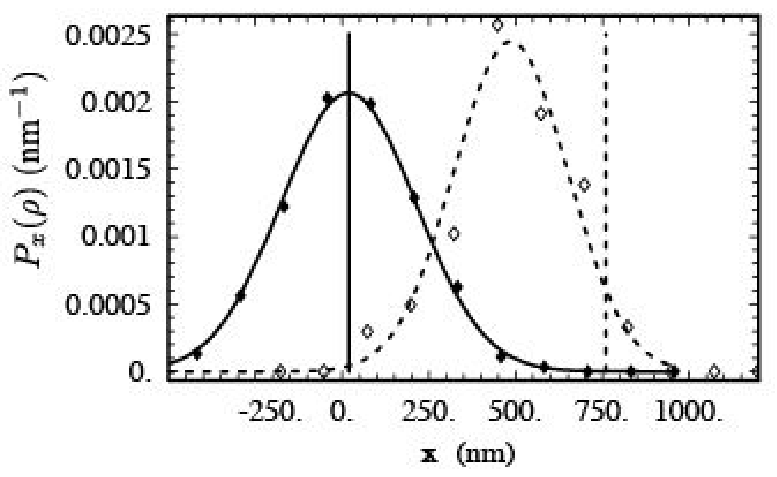}{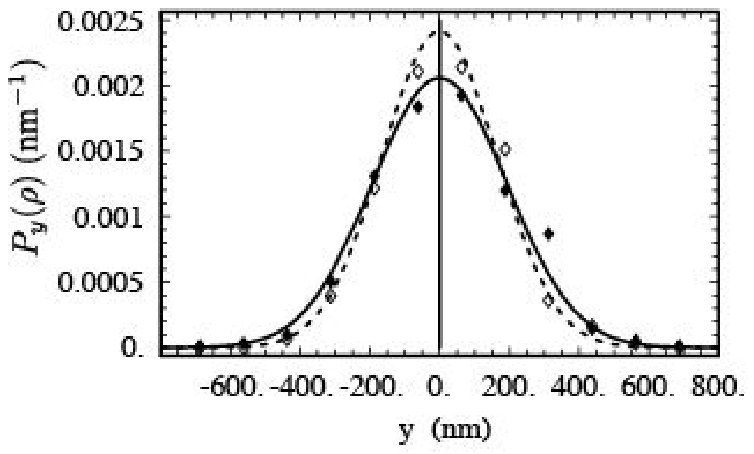}

The step-probability distributions (\fref{FIGHISTO3DPROJX}) obtained in this way show that at small $\rho'$ there is no preferred direction for the next time step; for larger $\rho'$, the tether is stretched and exerts a restoring force on the bead, so the step distribution shows a bias to diffuse toward the attachment point (the $-\hat\mathbf x$ direction). We now seek a convenient analytical representation of these distributions, both for computing the likelihood function $P\stot(\CO)$, and also for simulation purposes.

Each distribution is seen to be roughly a 2D Gaussian, with one principal axis along the radial direction to the attachment point. After rotating $\vr'$ to lie along the $\hat\mathbf x$ axis as described above, the principal axes of the distribution are the  $\hat\mathbf x$- and $\hat\mathbf y$-axes. The center point also lies on the $\hat\mathbf x$-axis, and is increasingly shifted from $\rho'\hat\mathbf x$ toward the anchor point $(0,0)$ as $\rho'$ increases. Accordingly, we can characterize $\transf\sun(\vr|\vr')$ for each fixed $\vr'=(\rho',0)$ by finding its mean $\exv x$ and the variances in the $x$ and $y$ directions.  The mean $\exv{y}$ equals zero (see for example \fref{FIGHISTO3DPROJX}), as it must by rotational invariance. Thus we seek three functions of $\rho'$ to characterize the histograms: the mean $\mu(\rho')\equiv\exv{x}_{\rho'}$, and the variances $\sigma^2_{x}(\rho')$ and $\sigma^2_{y}(\rho')$. Examples of these functions for illustrative values of $\rho'$ appear in \fref{MEANVARTRAIN}.

Again, the mean shift function $\mu$ reflects the drift of the bead toward the central point under the influence of a restoring force from the tether's entropic elasticity. As expected, it vanishes at $\rho'=0$ and becomes more negative with increasing $\rho'$. (The variances have weaker dependences on $\rho'$.) We constructed three interpolating functions to represent our results, which we took to be a 3rd-order polynomial for $\mu$ and sigmoids for $\sigma_{x,y}$ (see \fref{MEANVARTRAIN}):
\begin{equation}
\begin{array}{cc}
\mu_{ x}(\rho) = a_{0}+a_{1} \rho+a_{2} \rho^{2}+a_{3} \rho^{3}\\
\sigma^{2}_{x}(\rho) = {b_{0}}/(1+{\ex{(\rho-b_{1})/b_{2}}})+b_{3}\\
\sigma^{2}_{y}(\rho) = {c_{0}}/(1+{\ex{(\rho-c_{1})/c_{2}}})+c_{3}\\
\end{array}
\pnlabel{e:train}
\end{equation}

\ifigab{MEANVARTRAIN}{Empirical fit functions for the mean (\textit{left}) and variance (\textit{right}) of the 2D histograms (see e.g.\ \fref{FIGHISTO3D}) for experimental control data corresponding to unlooped (\textit{solid} symbols) and looped (\textit{open} symbols) tether states. Corresponding fit parameters for the functions of \eref{train} are located in \aref{b}.}{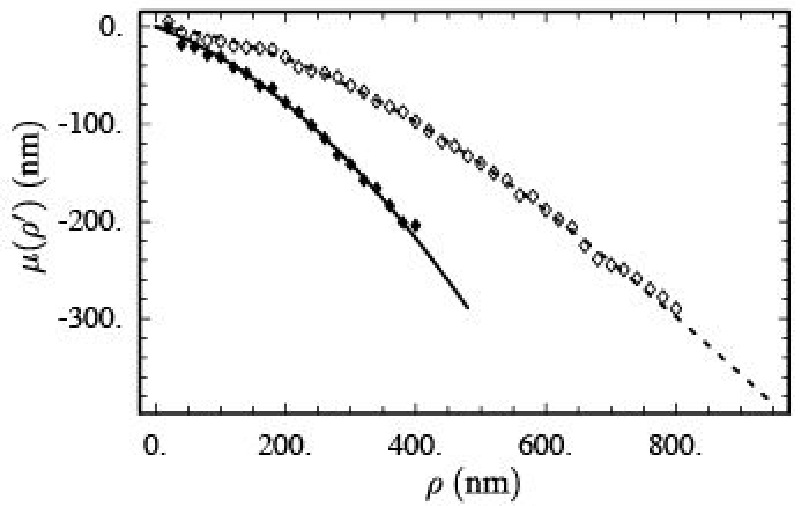}{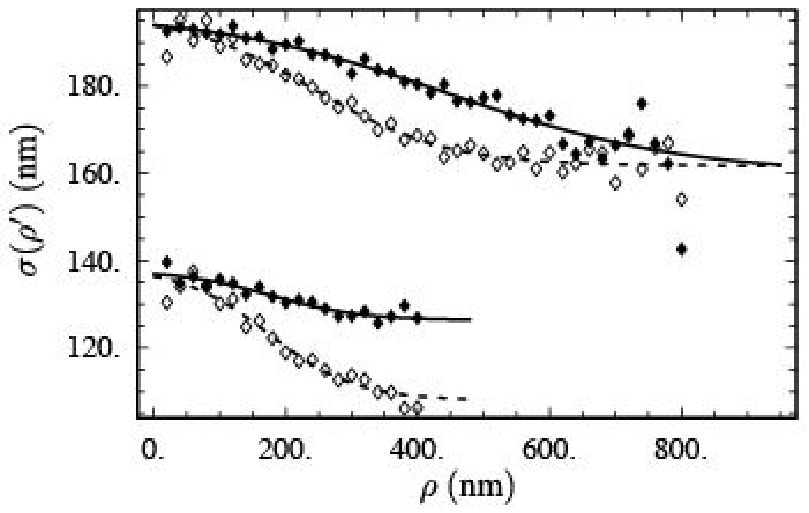}

Using these fit functions, we can represent the observed step probabilities starting from the point $\vr'=(\rho',0)$ as the product of 1D Gaussian distributions in $x$ and $y$:
\begin{eqnarray}
\transf\sun(\mathbf{r}|\mathbf{r}')&=&\gfunc_x( x|\rho')\cdot \gfunc_y( y|\rho')\mbox{
\ where}\pnlabel{e:pxy}\\
\gfunc_x( x|\rho')&=&(2 \pi \sigma^{2}_{x}(\rho'))^{-1/2} \exp \left(\frac{-( x-\mu(\rho'))^{2}}{2
\sigma^{2}_{x}(\rho')}\right)\pnlabel{e:px}\\
\gfunc_y( y|\rho')&=&(2 \pi \sigma^{2}_{y}(\rho'))^{-1/2}\exp\left(\frac{-y^{2}}{2 \sigma^{2}_{y}(\rho')}\right)\pnlabel{e:py}
\end{eqnarray}
For arbitrary $\vr'$ (not necessarily on the $\hat\mathbf x$ axis), we evaluate the probability by rotating $\vr'$ to the $\hat\mathbf x$-axis, rotating $\vr$ by the same amount, and evaluating \eref{pxy} on the components of the rotated $\vr$.

The procedure summarized in \erefs{pxy}--\ref{e:py}) is conceptually simple. However, we chose to improve the accuracy of our calculations by using a small elaboration. We noted that, like any Gaussian, the distribution defined above is nonzero for any $x$ and $y$. In reality, however, the DNA tether sets an absolute limit on $\rho$ beyond which the probability must be exactly zero. Not surprisingly, we found that following the procedure outlined above yielded simulated time series that occasionally violated this limit. Although the effect of this error may be minor for the unlooped step distribution, we reasoned that for the looped distribution it could interfere with looping state identification.

Accordingly, we modified our formula for $\transf\sun(\mathbf{r}|\mathbf{r}')$ to account for the limit in an approximate (and computationally inexpensive) way: We replaced \eref{px} by a truncated Gaussian function. That is, we took $\gfunc_x(x|\rho')$ to be zero for $x>\rhomax$, and a Gaussian with modified parameters for $x<\rhomax$. The modified parameters were chosen in such a way that the truncated Gaussian would again have the mean $\mu(\rho')$ and variance $\sigma^2_x(\rho')$ shown in \fref{MEANVARTRAIN}. That is, for each value of $\rho'$, the $\mu$ and $\sigma^{2}_{x}$ determined empirically from the data were not used directly; instead we found a new Gaussian, with modified parameters $\tilde{\mu}$ and $\tilde{\sigma}^{2}_{x}$, which has mean $\mu$ and variance $\sigma^2_x$ when the probability of values greater than $\rhomax$ is set to zero. Details of this transformation appear in \aref{b}.

\subsubsection{Optimization\pnlabel{s:o}}
Initially, the optimum lifetimes were found by evaluating $P\stot(\CO)$ on an evenly-spaced logarithmic grid of values for $\tlf$ and $\tlb$, and the maximum $P^{*}\stot(\CO)$ was found. The resulting likelihood surface is smooth (\fref{FIGLOGLIK}), so the peak likelihood can be determined more precisely by fitting a 2D quadratic in the neighborhood of the optimum lifetimes. The uncertainty of the optimum lifetimes corresponding to $\ln[P^{*}\stot(\CO)]-2$, i.e., enclosing 97\% of the probability, were estimated along the principal axes of this 2D quadratic to account for any correlation between the estimated lifetimes. In order to facilitate this iterative process, an automated simplex solver routine was implemented to find the maximum \cite{pres92a}.

\ifig{FIGLOGLIK}{Evaluation of $\ln[P^{*}\stot(\CO)]$ on a logarithmically-spaced grid of $\tlf, \tlb$ lifetimes corresponding to data from \fref{FIGTIMERHO}.}{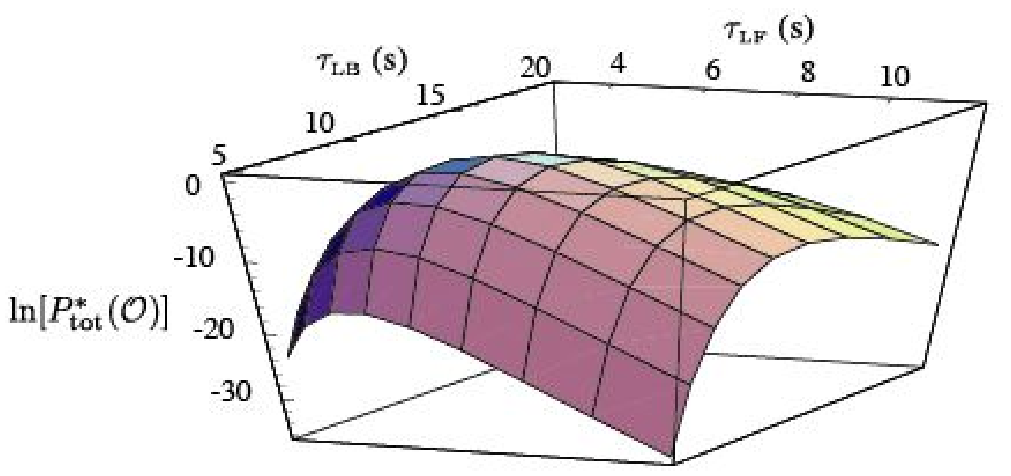}

\subsubsection{Simulation strategy\pnlabel{s:ss}}
Simulations of bead motion, with and without dynamic looping, were performed in \textsl{Mathematica} to test the DHMM model. Each step of the simulation (\textit{a})~first determined whether or not to remain in the current looped state, and then (\textit{b})~the next spatial position was determined appropriate for the particular loop state.

In more detail, (\textit{a})~If the initial state was looped, a pseudorandom number was used to determine whether to transition to the unlooped state with probability $\frac{\Delta t}{\tlb}$. If the initial state was unlooped, and if $\rho<\rhomax$, then a transition to the looped state was allowed with probability $\frac{\Delta t}{\tlf}$. Next (\textit{b})~a $(\Delta x, \Delta y)$ pair was drawn from the appropriate distribution obtained in \sref{osdf}. That is, $\Delta y$ was Gaussian distributed, and similarly for $\Delta x$ except that steps resulting in $\rho>\rhomax$ were discarded and the step repeated in order to achieve a truncated Gaussian as discussed earlier.

\section{Results and discussion\pnlabel{s:r}}
\Sref{i} outlined how we determined $\transf\sun(\mathbf{r}|\mathbf{r}')$ and $\transf\sloop(\mathbf{r}|\mathbf{r}')$  from the control data, then used these functions to simulate tethered particle motion. In this section we compare one-state simulations with control data and two-state simulations with dynamic looping data.  We also use DHMM to locate a change in looping dynamics, possibly indicating a change in protein occupancy of one or both of the \textit{lambda} operators.

\subsection{One-state modeling\pnlabel{s:oss}}
To validate our approach to equilibrium, tethered Brownian motion, we wrote a simple simulation using the step-distribution functions obtained from adjacent video frames (\sref{i}) and compared the resulting trajectories with the experimental control data. The resulting simulated time series $\{\rho_t\}$ are difficult to distinguish from actual data by inspection (not shown), so we compared the equilibrium properties of the motion using the radial probability distributions (\fref{FIGHISTOS}) and the dynamic properties using the autocorrelation function (\fref{FIGAUTO}). Both of these non-trivial checks agree fairly well with the actual data, although for unknown reasons the equilibrium distribution of the longer tether is captured better by the model than the shorter tether.

We also applied our two-state DHMM approach to permanently-unlooped experimental control data, to see if the algorithm would incorrectly report any looping transitions. Instead, the algorithm correctly reported loop-formation times that were proportional to the length of the data set (that is, consistent with infinity), and loop-breakdown times approaching the sampling interval (that is, consistent with zero). Simulations of unlooped motion behaved similarly. As expected, DHMM applied to permanently-looped experimental control data and simulations did not detect any false loop breakdown events with the trends for loop breakdown and formation lifetimes exchanged from the unlooped case. Another consistency check was to verify that the step distribution functions originally calculated from the experimental control data \sref{osdf} were the same for the simulations.

Finally, we found the maximum likelihood transition sequence corresponding to the optimal lifetime values \cite{rabi89a}. No false looping transitions were reported for either of the control data sets, indicating that our rate for false positives for loop formation and breakdown is low.

\subsection{Dynamic looping\pnlabel{s:dl}}
We applied our two-state DHMM algorithm to the part of the time series in \fref{FIGTIMERHO} that displays dynamic looping (the region between 650 and 1750 seconds). The algorithm reported $\tlf=5.8\,\sunit\pm25\%$ and $\tlb=9.9\,\sunit\pm25\%$; it also determined the most-likely sequence of state transitions (see Fig.~2 of \cite{beau07a}). To see whether our model really detects dynamic transitions between the two tether lengths in this parameter regime, we simulated a 20-minute dynamic looping data set with these same lifetime values, applied DHMM to the simulated data, and compared the reported lifetimes to those we had input to the model. We also know each of the transitions in the simulated data, so we could check the ability of our algorithm to reconstruct these. Indeed, applying DHMM to simulations yielded reported lifetimes that agree with these values within uncertainty: The agreement was within 20\% of the known answer (see \tref{mlres}). The maximum-likelihood state transition sequence corresponding to the optimum lifetimes successfully detects $\sim 85$--$90\%$ of the known transitions, with missing events usually less than $\sim 1\, \sunit$ in duration.%
\footnote{Note however that, like any Hidden Markov method, DHMM is not specifically sensitive to missing events. Indeed in DHMM, event identification is performed \textit{after} the lifetimes are determined, and thus does not affect the determination of the lifetime. Instead of binning identified events and fitting the resulting histogram, DHMM directly maximizes the likelihood that a function describes the entire data set. If the time scale of events is too short compared to other time scales in the problem, DHMM reports that fact via its estimates of errors in the fit parameters.}

\begin{table}
\caption{Comparison of results for three simulations, their average, and data (from \fref{FIGTIMERHO}).  The inferred lifetimes for the data were used as input for the simulations.  All trials had a total number of points $M = 29\,424$. The second number in the event column is the (known) number of events.  Lifetimes are in seconds and other terms are nondimensional.  \pnlabel{t:mlres}}

\begin{indented}
\lineup
\item[]\begin{tabular}{@{}*{6}{l}}
\br
$\0\0$&\# Events&$\tlf$&$\tlb$&$\ln[P\stot(\CO)]$&$\frac{1}{M}\ln[P\stot(\CO)]$\cr
\mr
Data        &\0\0 50 &5.8&$9.9$&$\0-374\,089$ &$\0-12.714$\cr
Sim. \#1    &\0 68/75&4.9&$9.2$&$\0-376\,951$ &$\0-12.811$\cr
Sim. \#2    &\0 57/64&6.4&$9.3$&$\0-378\,173$ &$\0-12.853$\cr
Sim. \#3    &\0 62/77&5.6&$8.3$&$\0-378\,228$ &$\0-12.855$\cr
Average Sim.&62.3/72 &5.6&$9.0$&$\0-377\,784$ &$\0-12.840$\cr
\br
\end{tabular}
\end{indented}
\end{table}

We also simulated looping with lifetimes different from those seen in the experimental data, see \aref{c}.  \resub{10 additional simulations, similar to those reported in \tref{mlres}, indicate that scatter in the lifetimes determined by DHMM is no worse than the minimum expected from variability due to the finite sample size (data not shown).} Simulations with longer lifetimes $\tlf=\tlb=20\, \sunit$ and a total time of $\approx60\,$minutes to give more transitions were similarly successful (agreement of lifetimes within 10\% and 92\% success in detecting events, data not shown).



\resub{As discussed in \sref{pw}, in both the threshold method and traditional HMM the time resolution depends on how the data are analyzed (i.e., window size and degree of thinning respectively). To demonstrate the robustness of the DHMM method, we subdivided our data set (taken at a frame rate of $(20\,\msunit)\inv$) into two subsets with $\Delta t=40\,\msunit$, and also into four subsets with $\Delta t=80\,\msunit$ and computed lifetimes for all of these subsets. All lifetimes agree within the uncertainty of the method (data not shown; note that \erefs{train}--\ref{e:py}) must be re-calculated for different $\Delta t$).}

\subsection{Detection of very long-lived state transitions\pnlabel{s:dvllst}}
So far, we have used the peak of the likelihood function $P\stot(\CO)$ and its vicinity to determine the optimum looping lifetimes and their uncertainty.  We can take further advantage of the likelihood function to assess the uniformity of the dynamics. Our motivation for undertaking this study is that in a real DNA-looping system there are certainly more than two discrete states: E.g., individual repressor proteins can bind and unbind to their operator sites \cite{wong07a}. Indeed, the data studied in this paper (from \cite{beau07a}) came from a system with two sets of three operators, leading to a large set of potential occupancy patterns. Presumably the obvious change in behavior in \fref{FIGTIMERHO} at $t=700\,\sunit$ reflects the arrival of another repressor at an operator site, enabling loop formation. But there seems also to be a less obvious transition in the data around $t=1250\,\sunit$, from one looping regime to another one with different kinetics. Can DHMM help us to locate such changes objectively?

We reasoned that if the data had uniform, two-state looping behavior for the entire recording, then the log-likelihood of the whole would be equal to the sum of the log-likelihoods of its parts, and also the best-fit lifetimes should come out roughly equal for each part separately as for the whole. To test this, we adapted a procedure used by \rref{watk05a}: We again restricted the data in \fref{FIGTIMERHO} to the region after $650\,\sunit$, but this time we also divided the data into two regions, $A$ from 1 to $M'$ and $B$ from $M'$ to $M$, and then separately applied DHMM on the two regions to determine \resub{$\ln{P\stot(\CO_{AB})}\equiv \ln{P\stot(\CO_{A})}+\ln{P\stot(\CO_{B})}$.} The resulting \resub{$\ln{P\stot(\CO_{AB})}$} has a single, very sharp peak at a particular value of $M'$ (\fref{FigLLpeak}). In agreement with visual inspection, the optimal value of $M'$ is roughly two-thirds of the way through the retained data, that is, around $t=1300\,\sunit$ in \fref{FIGTIMERHO}. Moreover the reported optimal lifetime values in the two subsets are quite different: $(\tlf, \tlb)_{A} \simeq (10\, \sunit, 20\, \sunit)$ and $(\tlf, \tlb)_{B} \simeq (3\, \sunit, 4\, \sunit)$. When we repeated the procedure with simulated data, with a transition at $M'=0.6M$ from looping with these $(\tlf, \tlb)_{A}$ and $(\tlf, \tlb)_{B}$ values, we found a peak in $\ln P\stot(\CO_{AB})$ that was very similar to the one found with the experimental data (\fref{FigLLpeak}). In contrast, when we applied the procedure to simulated constant, two-state dynamics, we found as expected that $\ln {P\stot(\CO_{AB})}$ was insensitive to $M'$.

\ifig{FigLLpeak}{See text.  The peak in $\ln{P\stot(\CO_{AB})}$ (\textit{solid line}) suggests that the data from \fref{FIGTIMERHO} is composed of two regions at $M' \simeq 61\%M$ with different kinetics. The \textit{gray/dotted} lines are the results from simulated data with/without a transition at $M' \simeq 61\%M$, respectively.  For comparison purposes, each likelihood has had its peak vertically shifted to 0 by subtracting a constant.}{FigLLpeak.eps}


\section{Conclusion and outlook\pnlabel{s:d}}
Our goal was to enhance the ability of tethered particle experiments to study DNA looping kinetics \textit{in vitro}. For illustration, we used a simplified 2-state model that includes the one looped and one unlooped tether state, i.e.\ we did not account for the various possible occupancies of the repressor protein on the DNA binding sites, but our method can readily be extended to include such details. Other DNA looping systems, e.g., \textit{lac}, could also be studied with DHMM, including multiple-state models to study the multiple length looped states recently observed~\cite{wong07a}.

Previously,  threshold methods were applied to the data to quantify the kinetic rates between various looped and unlooped states. This method involves filtering the data to extract the transitions from the noisy diffusive motion of the bead, then fitting a (single or double) exponential to the tail of a histogram of the dwell times. Both filtering and histogramming discard potentially useful information; moreover, at least in the system we studied, the choice of filter window can influence the reported results. Our DHMM method avoids any such steps.

We reasoned that hidden Markov modeling would be a good way to learn about the hidden conformation of the DNA tether from the observed motion of the bead, because the observed motion can be statistically quantified and simple models used to describe the unobserved state of the tether. We represent the uninteresting diffusive motion empirically from control experiments, and then model the looping assuming exponential kinetics with unknown parameters corresponding to the loop formation and breakdown lifetimes. Then we maximize the likelihood that the experimental data (or a simulation) came from our proposed model. Our method is implemented in a computationally efficient code, available in the online supplement to \cite{beau07a}. Since there is no filtering and no binning of the data in DHMM, the kinetic parameters can be determined unambiguously. If desired, the most-likely transition state sequence can also be determined \cite{beau07a}.

\resub{It is easy to obtain unlooped control data to train the DHMM algorithm by omitting repressor protein during the experiment; however looped control data can be more challenging.  For the \textit{lambda} system considered here, infrequent yet long-lived looped states made this a relatively simple task.  In other systems alternatives exist, including separate experiments on constructs with shorter length tethers corresponding to the expected looped length or, if available, mutant repressor proteins with stronger affinity for DNA that result in permanently looped tethers.  In our system, we verified the robustness of DHMM to the model of the looped state by varying the fit parameters in \eref{train} by $\pm 10\%$ and noting that $\tlb$ and $\tlf$ remained unchanged (data not shown).}

Another advantage of DHMM is its ability to, at least partially, compensate an experimental bias inherent in the tethered particle method: Loops cannot form between successive measurements if the DNA is in an extended conformation due to the time for the bead to diffuse to a location closer to the anchor point. This effect inflates the observed loop formation times relative to the case of interest (free DNA in solution); indeed, in our simulations, where the lifetimes are known \textit{a priori}, we found that this effect inflates $\tlf$ by $\sim30\%$. Our DHMM model allowed us to compensate for it, by allowing loops to form only when $\rho<\rhomax$.

The recent incorporation of single--particle tracking into the TPM method was essential for us, because it allows rapid and precise measurements of bead position that are required for DHMM. In particular, the ability to simultaneously track \textit{multiple} tethered beads is helpful for removing instrumental drift \cite{nels06a}. Future experiments could in principle remove drift from the data entirely by simultaneously tracking a fixed, fiducial marker object.

We applied DHMM to one illustrative experimental dataset of \textit{lambda} DNA with 200$\,\nMunit$ cI and determined loop formation and breakdown lifetimes of $\sim$6 and 10$\,\sunit$ respectively. One surprisingly biologically relevant application of this work is that, at physiological [cI] corresponding to the lysogenic state, the loop is not permanently closed. We also found it interesting that these lifetimes are neither representative of all the beads that were observed in \cite{beau07a} (data not shown), nor even for the entire observation time of any single bead (see \fref{FIGTIMERHO}). Prior to adding cI, most beads have nearly identical tethered diffusive motion; however, after addition of cI, the kinetics of looping varied widely. Some beads were mostly unlooped with occasional looping events, some beads were the inverse of this, others showed long periods of dynamic looping, and some like \fref{FIGTIMERHO} seemed to show very sharply-defined changes between long-lived (often $>10\,$min) regimes of homogeneous behavior. One hypothesis to explain these long-time looping trends is that the occupancy of cI protein among the 6 \textit{lambda} binding sites changes, resulting in periods with more or less stable loops. For example, when all 6 sites are occupied a very stable loop might form, whereas  4-sites occupancy could result in a less stable, but still detectable, loop. Future experiments with fewer operators, or perhaps fluorescent cI protein, could be used to test this hypothesis directly.

One technique for considering such complex kinetic scenarios is to use a more elaborate state diagram; however, we explored a different approach that might be appropriate if cI proteins are binding and unbinding on a time scale much slower than loop formation, as we indeed observed. We reasoned that in this case, DNA looping data could be adequately represented by a concatenation of 2-state models, each with different kinetics, rather than by a far more elaborate model with many states. There appear to be three such regions in the data shown in \fref{FIGTIMERHO}. We removed the first, mostly unlooped region, to focus on the faster dynamics of the later region, which we split in all possible ways into two subregions. We analyzed each subregion separately using DHMM, and found the partition that resulted in the highest total likelihood, which was also much higher than if the two regions were assumed to be one homogeneous kinetic regime. The sharp peak in \fref{FigLLpeak} indicates that DHMM is a sensitive tool to localize such subtle transitions in time.

\ack{We thank Laura Finzi, Chiara Zurla, Carlo Manzo, and David Dunlap for providing the data from \cite{beau07a} and for many useful discussions. We thank Rob Phillips for recommending HMM methods for this problem and Seth Blumberg, Lin Han, Randall Kamien, and Liam Paninski for useful discussions. This work was supported in part by the Human Frontier Science Programme and NSF grants DGE-0221664, DMR04-25780, and DMR-0404674.}

\section{Glossary and notation\pnlabel{s:abb}}
See \tref{notation} for mathematical symbols.

\begin{table}
\caption{Summary of notation in this paper.\pnlabel{t:notation}}

{\renewcommand{\baselinestretch}{1.3}\small
\begin{tabular}{@{}ll}
\br
$\vr\cut{_{\perp}}=(x,y)$&drift-subtracted, projected bead center position, relative to anchoring \cr
\,&point\\

$\rho,\rhomax$&length of $\vr\cut{_{\perp}}$, and its maximum observed value
\\
$z$&height of tethered particle
\\
$t,\Delta t$&time and time difference, measured in units of 40$\,\msunit$
\\
$M;\ M'<M$&total number of data points in a time series; and in a selected subset
\\
$\CO=\{\vr_{\cut{\perp,}t}\}$&time series of observed $\vr\cut{_{\perp}}$
\\
$q,\CQ=\{q_t\}$&discrete variable describing DNA looping state, and corresponding\cr
 \,&time series
\\
$P(\vr_{t},q_{t})$&joint probability of bead position and looping state at time $t$
\\
$\pi(\vr,q)$&same as $P(\vr_1,q_1)$
\\
$p\sbead(\mathbf{r}|q)$&pdf of bead position given looping state, neglecting diffusive \cr
\,&correlations in $\vr$
\\
$\transf(q|q')$ &looping state transition matrix, neglecting this distribution's\cr
\,&dependence on observed position
\\
$\tlf,\tlb$&time constants for loop formation and breakdown (defined in \erefs{dmathmm},\ref{e:dmatdhmm}))
\\
$\tau\sdiff$&diffusion time for bead to traverse its range of motion
\\
$W$&filter time constant used in windowing/threshold method (\sref{tca})
\\
$P\stot(\CO), P^*\stot(\CO) $&predicted likelihood for entire time series in a given model\eref{ptotdhmmMat},\cr
\,&and its optimum value
\\
$\transf\sun(\vr|\vr'), \transf\sloop(\vr|\vr')$&step probability distributions for unlooped and permanently looped\cr \,&tethers respectively
\\
$\Dmat\shmm(\rho),\ \Dmat\sdhmm(\vr|\vr')$&$2\times2$ matrices defined in \erefs{dmathmm}, \ref{e:dmatdhmm})
\\
$\mu_x, \sigma_x^{2},\sigma_y^{2}$&functions of $\rho$ characterizing
$\transf(\vr|\vr')$ (see \eref{train})\\
$\gfunc_{x}(x|\rho),\ \gfunc_{y}(y|\rho)$&gaussian distributions
chosen to represent $\transf(\vr|\vr')$ (see \eref{pxy})\\
\br
\end{tabular}
}
\end{table}

Tethered Particle Motion (TPM): Class of biophysical experiments that infer the properties of a polymer tether, such as DNA, by tracking the position of a $\mu$m-size bead that is attached to one end of the tether.

DNA Looping: protein-stabilized loops of DNA that form and break when protein binds to specific sites along the DNA in vivo and in vitro.

Diffusive Hidden Markov Model (DHMM): a modification of traditional hidden Markov methods that accounts for the diffusive properties of the TPM particle to determine kinetic rates for DNA loop formation and breakdown in single molecule experiments of DNA looping.

Threshold Method: an analysis method for determining kinetic rate constants from a signal by identifying transitions between states, and then fitting the dwell times between these state transitions to a model (such as an exponential).

Window: The width in time of the filter applied to a noisy time trace in order to elucidate the states for identification in the Threshold method.

\newpage
\appendix
\section{Bead diffusion in standard HMM\pnlabel{a:a}}
\ifig{FIGHMMPLATEAU}{Hidden Markov analysis of the data from \fref{FIGTIMERHO} illustrate how the diffusive time scale of the bead and tether dominate the inferred lifetimes when the measurement sampling interval is faster than the time scale for bead/tether diffusion. If the sampling time is artificially slowed down by thinning the data then the inferred lifetimes plateau approximately to the values obtained by DHMM (\textit{solid/open} symbols refer to the inferred lifetimes for loop formation/breakdwon).  For a thinning of '2' all of the odd points are separated from the even and then these two data sets are concatenated to reconstitute the original length trace.  An analogous process was applied for higher order thinning.  We verified (data not shown) that HMM analysis of the concatenated recording gave results approximately equal to the average of the results if each thinned trace was considered individually.} {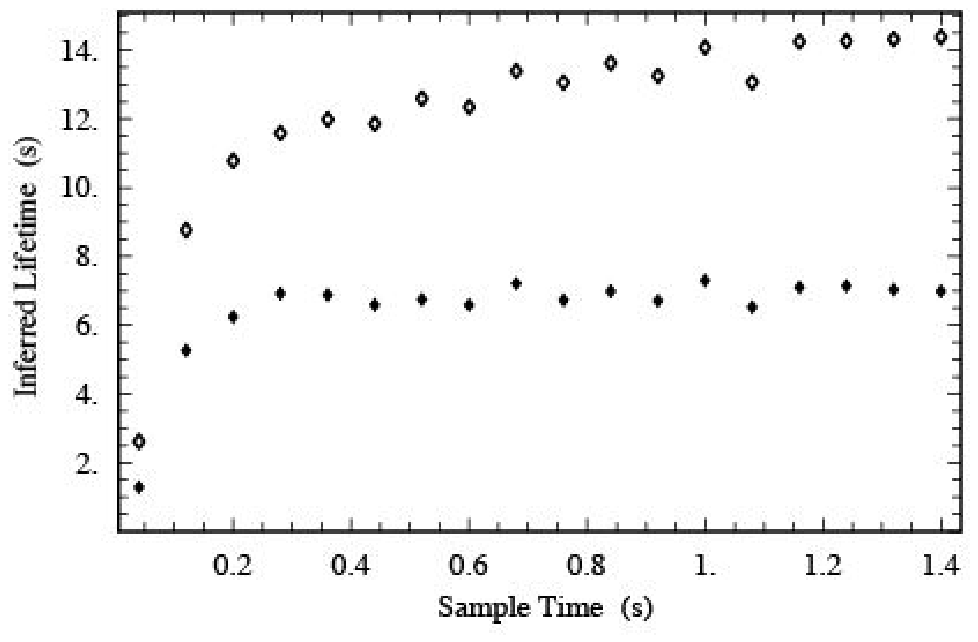}

As discussed in \sref{tha}, applying the standard hidden Markov analysis \eref{ptothmm} to tethered particle experiments resulted in unrealistically short lifetimes for loop formation and breakdown that are on the same order as $\tau\sdiff$.  The most probable looping sequence for $\CQ $ corresponding to these fast lifetimes can be calculated (by a method described in \cite{rabi89a}), and is consistent with the bead moving diffusively between regions of large and small $\rho$, effectively masquerading as very fast looping events. That is, the spurious reported transitions reflect the fact that successive video frames are not really independent measurements of the underlying tether state.

One can reduce this problem by thinning the data, thereby decreasing the influence of diffusion and making successive points more independent of each other. Repeating the HMM analysis, we find that the reported lifetimes increase. If this thinning process is repeated, the calculated lifetimes eventually plateau to a value roughly consistent with the time scale of the looping events as identified by eye (\fref{FIGHMMPLATEAU}). The looping sequence corresponding to these longer lifetimes also agrees with the looping events observed by eye in the raw data.

\section{Implementation details\pnlabel{a:b}}
\resub{\Sref{osdf} describes how we summarized empirical data, like those in \fref{FIGHISTO3D}, with a step-distribution function $\gfunc(x,y|\rho')$. Ideally $\gfunc$ should be chosen to be a function that vanishes when $x^2+y^2$ exceeds $(\rhomax)^2$, and falls smoothly to zero as that boundary is approached. To make our calculations tractable, we instead simplified by taking $\gfunc$ to be the product of a cutoff, shifted, 1D-Gaussian in $x$ times an ordinary Gaussian in $y$. Examination of many graphs like \fref{FIGHISTO3D} convinced us that this simplification adequately represented our empirical histograms. Moreover, we reasoned that, since we rotated axes to make the initial position lie along the $x$-axis, an extra excursion along $x$ was more likely to violate the tether condition than one along $y$. Perhaps most important, we checked that small changes to our choice of the empirical function $\gfunc$ have little effect (see \sref{d}). }

To implement efficient calculation of the truncated Gaussian $\gfunc_x(x|\rho')$, we evaluate a look--up table for the Gaussian with mean, variance, and normalization ($\tilde{\mu}(\rho'),
\tilde{\sigma}^{2}(\rho')$ and $\tilde{N})$ such that when $\rho>\rhomax$ this Gaussian is zero and satisfies the mean and variance $(\mu(\rho')$ and $\sigma_{x}^2(\rho'))$ determined empirically from
data in \cite{beau07a}:

\begin{equation}
\int_{-\infty}^{\rhomax}\dd x \frac{1}{\tilde{N}}\ex{
{-(x-\tilde{\mu})^2}/({2 \tilde{\sigma}^{2}})}=1,\quad
\int_{-\infty}^{\rhomax}\dd x \frac{x}{\tilde{N}}\ex{
{-(x-\tilde{\mu})^2}/({2 \tilde{\sigma}^{2}})}=\mu(\rho')
\end{equation}\begin{equation}
\int_{-\infty}^{\rhomax} \dd x\frac{x^2}{\tilde{N}}\ex{
{-(x-\tilde{\mu})^2}/({2 \tilde{\sigma}^{2}})}=\sigma_{x}^2(\rho')\\
\pnlabel{e:truncG}
\end{equation}

Such look-up tables were evaluated at 100 values of $\rho'$ for both (unlooped, looped) tether states using \eref{train} and parameters: a0=(0,0), a1= (-0.068, -0.238), a2= (-5.0e-4, -7.9e-4), a3 = (1.52e-7, 6.30e-8), b0= (35.1,30.6), b1=(161.75,107.95), b2=(242.3,173.8), b3=(100.16,66.42), c0=(37.11, 11.43), c1=(159.8,126.3), c2=(444.88, 177.06), c3=(180.72,67.86), where $\rho$ is measured in $\nmunit$.

\section{DHMM simulations\pnlabel{a:c}}
\resub{We set up a $5\times5$ log-spaced grid of different combinations of lifetimes centered on the optimum lifetimes we obtained from the experimental data. Then we generated simulated data sets using those assumed lifetimes, and applied DHMM, to see how well it could extract them (\fref{FIGGRIDTAU}). The agreement over the range ($1/4\times$ to $4\times$) was good to within the error bars estimated by the curvature of the log-likelihood function (\fref{FIGLOGLIK}). These error bars in part reflect the finite sample size $M$, which is often limited by experimental considerations. To check that last assertion, we generated 10 simulated data sets all with the same ``true'' lifetimes and of the same length $M$ as the experimental data discussed in the paper. Then we found the variation in the lifetimes deduced knowing the transition times, which represents the minimal variation, due to finite sample size. Finally we calculated the best-fit lifetimes from the simulated data using DHMM, and found the variation was similar to this minimal value (data not shown).}

\ifig{FIGGRIDTAU}{\resub{25 simulations were performed with input lifetimes $\tlf$ and $\tlb$ distributed on a log-spaced grid \textit{(gray crosses)} centered at the optimum lifetimes that we obtained in the text from experimental data. Each simulated data set had length $M$ equal to that of the experimental data. Next, DHMM was performed for each simulated data set, to determine the most-likely lifetimes \textit{(black symbols)}.}} {figGridtau.eps}

\clearpage


\bibliographystyle{unsrt}
\section*{References}
\bibliography{hmm}

\end{document}

\newpage
\centerline{axes labels}
\penumerate{\small
\item \fref{FIGTIMERHO}      x: Time\ (\sunit);        y: $\rho\ (\nmunit)$
\item \fref{FIGTIMEXY}      x1: Time\ (\sunit);       y1: $x\ (\nmunit)$
\item \fref{FIGTIMEXY}      x2: Time\ (\sunit);       y2: $y\ (\nmunit)$
\item \fref{FIGHISTOS}       x: $\rho\ (\nmunit)$;     y: $p_{\rm bead}(\rho)\ (\nmunit^{-1})$
\item \fref{FIGAUTO}         x: $\tau\ (\sunit)$;      y: $\ln\bigl(\half\langle\vr_{t}\cdot\vr_{t+\tau}\rangle\bigr)$
\item \fref{FIGHISTO3DPROJX}x1: $\rho\ (\nmunit)$;    y1: $P_{x}(\rho)\ (\nmunit^{-1})$
                            x2: $\rho\ (\nmunit)$;    y2: $P_{y}(\rho)\ (\nmunit^{-1})$
\item \fref{MEANVARTRAIN}   x1: $\rho\ (\nmunit)$;    y1: $\mu(\rho')\ (\nmunit)$
                            x2: $\rho\ (\nmunit)$;    y2: $\sigma(\rho')\ (\nmunit)$
\item \fref{FIGLOGLIK}       x: $\tlb$\ (\sunit);      y: $\tlf$\ (\sunit);   z: $\ln[P^{*}\stot(\CO)]$
\item \fref{FigLLpeak}       x: $M'/M$;                y: $P\stot(\CO_{AB})$
\item \fref{FIGHMMPLATEAU}   x: Sample Time\ (\sunit); y: Inferred Lifetimes\ (\sunit)
\item \fref{FIGMULTSIMS}     x: $\tlb$\ (\sunit);      y: $\tlf$\ (\sunit)
\item \fref{FIGGRIDTAU}      x: $\tlb$\ (\sunit);      y: $\tlf$\ (\sunit)

}\xpenumerate